\documentstyle[twoside,fleqn,espcrc2,epsfig]{article}
\newcommand{\AmS}{{\protect\the\textfont2
  A\kern-.1667em\lower.5ex\hbox{M}\kern-.125emS}}

\def\be{\begin{equation}}
\def\ee{\end{equation}}
\def\bea{\begin{eqnarray}}
\def\eea{\end{eqnarray}}

\title{\vskip-3.cm
{\baselineskip14pt
\centerline{\normalsize\hfill IFT-26/96, October 1996}
\centerline{\normalsize\hfill hep-ph/9612345}}
\vskip22.cm
{\baselineskip10pt
\begin{flushleft}
{\small\em To appear in the Proceedings of the 
Fourth International Workshop on Tau Lepton Physics, 16-19 September
1996, Estes Park, Colorado, U.S.A.}
\end{flushleft}}
\vskip-21.2cm
QCD tests in $\tau$ decays with optimized perturbation 
expansion}

\author{P. A. R\c{a}czka\address{Institute of Theoretical Physics,
Warsaw University,\\  
ul.\ Ho\.{z}a~69, PL-00-681 Warsaw, Poland}\thanks{Address after
October~1, 1996: Centre for Particle Theory, Department of Physics,
University of Durham, South  Road, Durham DH1~3LE, UK.}}
\begin{document}
\begin{abstract}
The next-to-next-to-leading order perturbative QCD corrections to
$R_{\tau}$ and the higher moments of the invariant mass distribution
in the hadronic tau decays are considered.  The renormalization scheme
dependence of these corrections is discussed.  The optimized
predictions are obtained, using the principle of minimal sensitivity
as a guide to select the preferred renormalization scheme.  A
simplified fit is performed, using $R_{\tau}$ and $R_{\tau}^{12}$, to
see how the use of the optimized expansion may affect the
determination of the $\alpha_{s}$ and the dimension six condensates
from the experimental data.
\end{abstract}

\maketitle
Recently there has been considerable interest in the QCD predictions
for the total hadronic width of the $\tau$ and the higher order
moments of the invariant mass distribution in the hadronic $\tau$
decays~\cite{ledepi-kl,aleph93,aleph95,cleo}. These quantities have
been used to 
obtain tight 
experimental constraints on $\alpha_{s}$ and the condensates.  In
spite of the high precision which may be obtained in fits to
the experimental data it is important to investigate in detail how the
results may be affected by the renormalization scheme (RS) dependence
of the perturbative QCD predictions. Usually the $\overline{MS}$
scheme~\cite{msb} is used to evaluate the perturbative QCD
corrections. However, there is a two-parameter freedom in the choice
of the RS in the next-to-next-to-leading order (NNLO), and there is no
{\em a priori} theoretical or phenomenological reason why the
$\overline{MS}$ scheme should be preferred.  The difference in
predictions obtained in various schemes is formally of higher order in
the coupling, but numerically it may be significant, particularly at
the energy scale of $m_{\tau}=1.777~\mbox{GeV}$.

The QCD prediction for the $R_{\tau}$ ratio 
\be
R_{\tau}=\frac{\Gamma(\tau \rightarrow \nu_{\tau} + hadrons)}
{\Gamma(\tau \rightarrow \nu_{\tau} e^{-} \overline{\nu}_{e})},
\ee
has the form~\cite{branapi}
\be
R_{\tau}=3S_{CKM}S_{EW}(1+
\delta^{tot}_{pt}+\delta^{tot}_{m}+\delta^{tot}_{SVZ}),
\ee
where $S_{CKM}=(|V_{ud}|^{2}+|V_{us}|^{2})\approx1$. The
factor $S_{EW}=1.0194$ represents the corrections from electroweak 
interactions.  The 
$\delta_{pt}^{tot}$ contribution denotes the purely perturbative QCD
correction, evaluated for three massless quarks. The $\delta_{m}^{tot}$
contribution denotes the correction from quark masses
($\delta^{tot}_{m}\approx0.009$). 
In the case of the 
 $R_{\tau}^{kl}$ moments of the
 invariant mass distribution $d\Gamma_{ud}/ds$ of the Cabbibo allowed
decays, which are defined by the
 relation~\cite{ledepi-kl}:
\be
R^{kl}_{\tau}=\frac{1}{\Gamma_{e}} \int_{0}^{m_{\tau}^{2}}
ds\,\left(1-\frac{s}{m_{\tau}^{2}}\right)^{k}
\left(\frac{s}{m_{\tau}^{2}}\right)^{l}
\frac{d\Gamma_{ud}}{ds},
\ee
where $\Gamma_{e}$ is the electronic width of $\tau$, the QCD
prediction has the form:
\be
R_{\tau}^{kl}=3\,|V_{ud}|^{2}\,S_{EW}\,
R^{kl}_{0}(1+\delta^{kl}_{pt}+\delta^{kl}_{SVZ}).
\label{rtkl}
\ee
$R^{kl}_{0}$ in~(\ref{rtkl}) denotes the parton model prediction. 
(The $\delta_{m}$ contribution is negligible in the case of
$R_{\tau}^{kl}$.) The $\delta_{SVZ}$ contributions in $R_{\tau}$ and
$R_{\tau}^{kl}$ denote nonperturbative QCD corrections calculated
using the $SVZ$ approach~\cite{svz}
\be
\delta_{SVZ}=\sum_{D=4,6...} c_{D} 
\frac{O_{D}}{m^{D}_{\tau}}.
\label{svz}
\ee
The parameters $O_{D}$ in Eq.(\ref{svz}) denote vacuum expectation
 values of the gauge invariant operators of dimension $D$.  The
 $c_{D}$ coefficients are in principle power series in the strong
 coupling constant, which depend on the considered quantity.

The contribution from the $D=4$ term in the SVZ expansion for
$R_{\tau}$ may be reliably expected to be small~\cite{branapi}, since
$O_{4}$ is well constrained by the sum rules phenomenology, and the
relevant coefficient function starts at $O(\alpha_{s}^{2})$.  However,
the $D=6$ contributions to $R_{\tau}$ is not suppressed, and there is
little information on the value of $O_{6}$. It was therefore
proposed~\cite{ledepi-kl} to treat $O_{D}$ as free parameters, which
are to be extracted together with $\alpha_{s}$ from a fit to the
experimental data for $R_{\tau}$ and the higher moments of the
invariant mass distribution. 

The analysis reported in~\cite{ledepi-kl,aleph93,aleph95,cleo}
involved the $R_{\tau}^{kl}$ 
moments with $(k,l)$ equal to $(1,0)$, $(1,1)$, $(1,2)$ and
$(1,3)$. However, if we are interested primarily in the possible
effect of RS dependence, we may simplify the discussion by considering
only the $R^{12}_{\tau}$ moment, for which --- similarly as for the 
$R_{\tau}$ --- the $D=4$ contribution is suppressed and there is
significant contribution from the $D=6$ term. Retaining in the SVZ
expansion only the $D=6$ term, which appears to be a dominant source
of the uncertainty in the nonperturbative sector, we obtain a simplest
set of the QCD predictions which allows for a self-consistent
extraction of $\alpha_{s}$ and $O_{6}$ from tau decays.

The perturbative QCD corrections $\delta_{pt}^{tot}$ and
$\delta_{pt}^{kl}$ may be expressed as a contour integral in the
complex energy plane~\cite{contour}, with the so called Adler
function~\cite{adler} under the integral. (Actually
$\delta_{pt}^{tot}=\delta_{pt}^{00}$.)  We have:
\be
\delta_{pt}^{kl}=\frac{i}{\pi}\int_{C}\frac{d\sigma}{\sigma}
f^{kl}(\frac{\sigma}{m_{\tau}^{2}})\delta_{D,V}(-\sigma),
\label{eq:cont}
\ee
 where $C\,$ is a contour running clockwise from
$\sigma=m^{2}_{\tau}-i{\epsilon}$ to $\sigma=m^{2}_{\tau}+i{\epsilon}$
away from the region of small $|\sigma|$. In the actual calculation we
assume that $C$ is a circle $|\sigma|=m_{\tau}^{2}$. The Adler
function is defined by the relation:
\be
(-12\pi^{2}) \sigma \frac{d\,}{d\sigma}
\Pi_{V}^{(1)}(\sigma)=3S_{CKM}[1+\delta_{D,V}(-\sigma)]
\ee
where $\Pi_{V}^{(1)}$ denotes the transverse part of the vector current
correlator. The function $f^{12}(\sigma/m_{\tau}^{2})$ has the form: 
\bea
f^{12}(x)&=&\frac{1}{2}-\frac{70}{13}x^{3}+
\frac{105}{26}x^{4} \nonumber \\
         & &+\frac{126}{13}x^{5}-\frac{175}{13}x^{6}+\frac{60}{13}x^{7}.
\eea
The function $f^{00}(\sigma/m_{\tau}^{2})$ has the form:
\bea
f^{00}(x)&=&\frac{1}{2}-x+x^{3}-\frac{1}{2}\,x^{4}.
\eea
The NNLO renormalization group improved perturbative expansion for
$\delta_{D,V}$ may be written in the form:
\begin{equation}
\delta_{D,V}^{(2)}(-\sigma) = a(-\sigma)[1+
r_{1}a(-\sigma)+r_{2}a^{2}(-\sigma)],
\label{del}
\end{equation}
 where $a=\alpha_{s}/\pi=g^{2}/(4 \pi^{2})$ denotes the running
coupling constant that satisfies the NNLO renormalization group
equation:
\begin{equation}
\sigma \frac{d\,a}{d\sigma} = - \frac{b}{2}
\,a^{2}\,(1 + c_{1}a + c_{2}a^{2}\,).
\label{rge}
\end{equation}
In the $\overline{MS}$ scheme we have 
$r_{1}^{\overline{MS}}=1.63982$ and~\cite{r1r2}
$r_{2}^{\overline{MS}}=6.37101$. 
The renormalization group coefficients for $n_{f}=3$ are $b=4.5$,
$c_{1}=16/9$ and $c_{2}^{\overline{MS}}=3863/864\approx4.471$.

If the Adler function is expanded in terms of $a(m_{\tau}^{2})$, then
the 
$\sigma$ dependence appears through the powers of
$\ln(-\sigma/m_{\tau}^{2})$. The contour integration is then
straightforward and the conventional NNLO expansion of
$\delta^{kl}_{pt}$ in terms of $a(m_{\tau}^{2})$ is easily obtained.
However, it was observed in~\cite{pivo92,ledepi92a} that one may also
keep under the 
integral the renormalization group improved expression for the Adler
function. In this case the contour integral has to be evaluated
numerically.  This results in the  essential improvement of the
conventional expansion, corresponding to the
all-order resummation of some of the corrections arising from analytic
continuation from spacelike to timelike momenta.

The QCD predictions calculated in the next-to-next-to-leading order
(NNLO) approximation with massless quarks depend on two RS parameters,
which in principle may be arbitrary. The two-parameter freedom in NNLO
arises because in each order of perturbation expansion we are free to
choose independently the finite parts of the coupling constant
renormalization constant. Different choices of the finite parts of 
the renormalization constants result in different definitions of the
coupling constant, which are related by finite renormalization. (The
formulas describing how the redefinition of the coupling affects the
coefficients $r_{i}$ and $c_{2}$ are collected for example
in~\cite{par-rt1a}.) Also the dimensional QCD parameter $\Lambda$
depends on 
the choice of the RS. In the NNLO there exists however a RS invariant
combination of the expansion coefficients~\cite{pms,dhar}:
\be
\rho_{2}=c_{2}+r_{2}-c_{1}r_{1}-r_{1}^{2}.
\label{rho2}
\ee
For the Adler function we have $\rho_{2}=5.23783$. 

The change in the expansion coefficients and the change in the
coupling constant compensate each other, but of course in the finite
order of perturbation expansion such compensation may only be
approximate, which 
results in the numerical RS dependence of the perturbative
predictions.  There has been intensive discussion on the prescriptions
for making an optimal choice of the RS~\cite{rsdep}. One of the most
attractive propostions is the choice based on the so called principle
of minimal sensitivity (PMS)~\cite{pms}, which singles out the scheme
parameters for which the finite order prediction is least sensitive to
the change of RS, similarly to what we expect from the actual physical
quantity.  (It should be emphasized that in our case the algebraic PMS
optimization equations~\cite{pms} do not apply and a nontrivial
numerical analysis of the perturbative prediction is required.)

The PMS optimization has to some extent a heuristic character and
therefore it is important to investigate the stability of the
predictions also with respect to non-infinitesimal changes of the
scheme parameters. This may be done by calculating the variation of
the predictions over a set of {\em a priori} acceptable schemes.  A
condition for selecting a class of acceptable schemes in NNLO has been
proposed in~\cite{par1}:
\be
|c_{2}|+|r_{2}|+c_{1}|r_{1}|+r_{1}^{2}\leq l\,|\rho_{2}|.
\label{allow}
\ee
 This condition is based on the observation, that the schemes with
unnaturally large expansion coefficients would give rise to extensive
cancellations in the expression for the RS invariant $\rho_{2}$.  The
constant $l$ controls the degree of cancellation in $\rho_{2}$ that we
want to allow. In particular, for the conventional QCD expansion the
PMS scheme lies approximately at the boundary of the region
corresponding to $l=2$~\cite{par1}.

A detailed discussion of the RS dependence of $\delta_{pt}^{tot}$ has
been presented in~\cite{par-rt} and would not be repeated here. An
important conclusion from~\cite{par-rt} is that the contour integral
resummation of higher order analytic continuation corrections  is very
important for 
ensuring the RS stability of the predictions. (This has been also
discussed in~\cite{ledepi92a}. The instability of the
conventional expansion for $\delta_{pt}^{tot}$ has been discussed
in~\cite{par-rt1a,par-rt1b,ledepi92a}.) Below we concentrate on the RS
dependence of 
$\delta_{pt}^{12}$.

\begin{figure}[htb]
 ~\epsfig{file=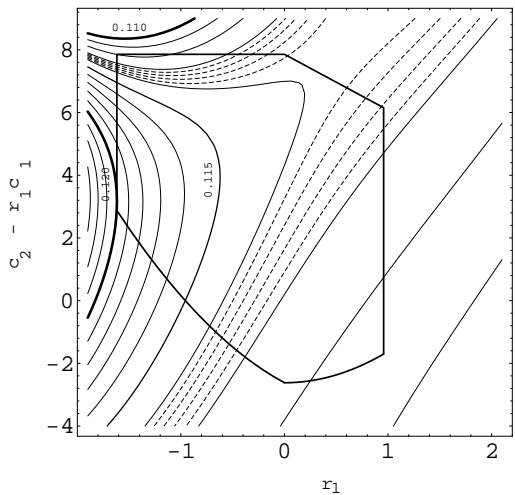,bbllx=209pt,bblly=303pt,bburx=403pt,bbury=488pt} 
\caption{The contour plot of $\delta_{pt}^{12}$ as a function of the
scheme parameters $r_{1}$ and $c_{2}$, for
$\Lambda_{\overline{MS}}^{(3)}=325~\mbox{MeV}$. For technical reasons
we use $c_{2}-c_{1}r_{1}$ on the vertical axis instead of $c_{2}$. The
boundary of the region of scheme parameters satisfying the
Eq.~\protect\ref{allow} is also indicated.}
\label{fig:cplot}
\end{figure}

The dependence of $\delta_{pt}^{12}$ on the scheme parameters is
illustrated in figure~\ref{fig:cplot} for
$\Lambda_{\overline{MS}}^{(3)}=325~\mbox{MeV}$.  (Similarly as in the
previous work~\cite{par-rt} we parametrize the freedom of choice of
the RS by the parameters $r_{1}$ and $c_{2}$, but for technical
reasons we use $c_{2}-c_{1}r_{1}$ on the vertical axis of
figure~\ref{fig:cplot} instead of $c_{2}$.)  The saddle point on
figure~\ref{fig:cplot} represents the PMS parameters. It is
interesting, that even though the expression for $\delta_{pt}^{12}$
has essentially non-polynomial character, the PMS parameters are well
approximated by $r_{1}=0$ and $c_{2}=1.5\rho_{2}$. We shall choose
these values as our optimized parameters. (The exact PMS parameters have
some dependence on the value of $\Lambda_{\overline{MS}}$. Also, for
very large values of $\Lambda_{\overline{MS}}$ the pattern of RS
dependence is more complicated than this shown in
figure~\ref{fig:cplot}. However, for all values of
$\Lambda_{\overline{MS}}$ 
the RS dependence in the vicinity of $r_{1}=0$ and $c_{2}=1.5\rho_{2}$
is very small.)  The set of scheme parameters, that involve the same ---
or smaller 
--- degree of cancellation in $\rho_{2}$ than our preferred parameters,
satisfies the
condition~(\ref{allow}) with $l=2$. The boundary of this set is indicated
on the figure~\ref{fig:cplot}. By calculating the variation of the
predictions over this set of scheme parameters we may estimate in a
quantitative 
way the sensitivity of the NNLO predictions to the change of RS. 
Thus obtained estimate may then  be compared
with a similar estimate for other quantities, for which the NNLO
predictions are known. It should be noted that although the $\overline{MS}$
parameters lie outside of the ``allowed'' region shown in the
figure~\ref{fig:cplot}, the numerical value of the
prediction in the $\overline{MS}$ scheme is close to the lowest value
attained in the ``allowed'' region.

It is interesting to note, that the RS dependence pattern of
$\delta_{pt}^{12}$ is quite similar to the pattern found in~\cite{par-rt}
for $\delta_{pt}^{tot}$, despite the fact that the conventional
expansions for these quantities appear to be quite different. Indeed,
in the conventional expansion  for $\delta_{pt}^{tot}$ we have
$r_{1}^{\overline{MS}}=5.2023$ and $r_{2}^{\overline{MS}}=26.3659$,
which gives $\rho_{2}=-5.4757$, while for $\delta_{pt}^{12}$ we have
$r_{1}^{\overline{MS}}=3.5795$ and $r_{2}^{\overline{MS}}=4.3441$,
which gives $\rho_{2}=-10.3614$. This seems to suggest that the
improved predictions obtained with contour integral expressions are
more natural, reflecting the 
common origin of the two corrections in a better way.

\begin{figure}[htb]
 ~\epsfig{file=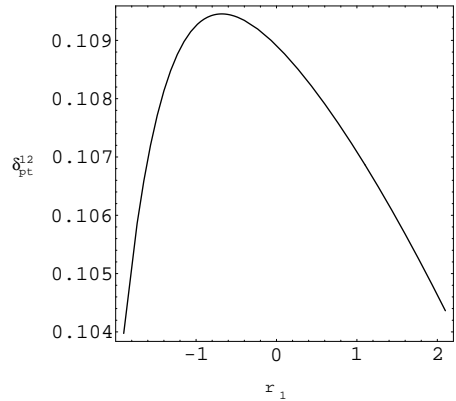,bbllx=223pt,bblly=318pt,bburx=389pt,bbury=474pt} 
\caption{NLO prediction for $\delta_{pt}^{12}$ as a function of the
scheme parameter $r_{1}$, for
$\Lambda_{\overline{MS}}^{(3)}=325~\mbox{MeV}$.}
\label{fig:nlo}
\end{figure}

In figure~\ref{fig:nlo} we show the RS dependence of the
next-to-leading order (NLO) prediction for $\delta_{pt}^{12}$. By numerical
optimization we find that in NLO $r_{1}^{PMS}\approx-0.64$. (Again,
this depends to some extent on $\Lambda_{\overline{MS}}$, but this
dependence has negligible effect on the numerical value of the
prediction.)

\begin{figure}[htb]
\center
 ~\epsfig{file=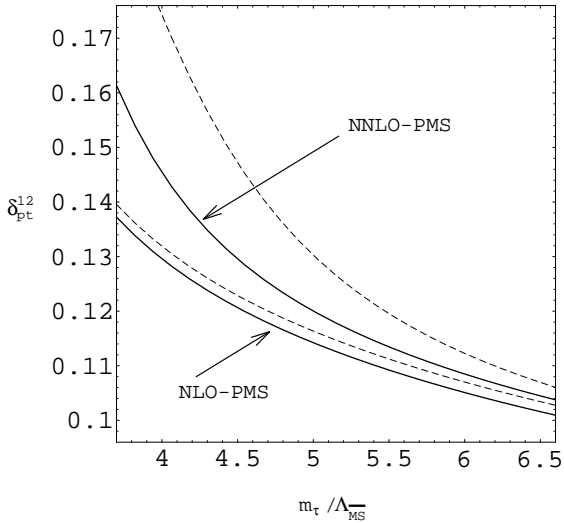,bbllx=202pt,bblly=295pt,bburx=410pt,bbury=497pt} 
\caption{The optimized predictions for $\delta_{pt}^{12}$, as a function of
$m_{\tau}/\Lambda_{\overline{MS}}^{(3)}$, obtained in NNLO (upper
solid curve) and NLO (lower solid curve).  The dashed lines indicate
variation of the predictions when the scheme parameters are changed
within the region satisfying the Eq.~\protect\ref{allow}.}
\label{fig:endep}
\end{figure}

\begin{figure}[htb]
\center
 ~\epsfig{file=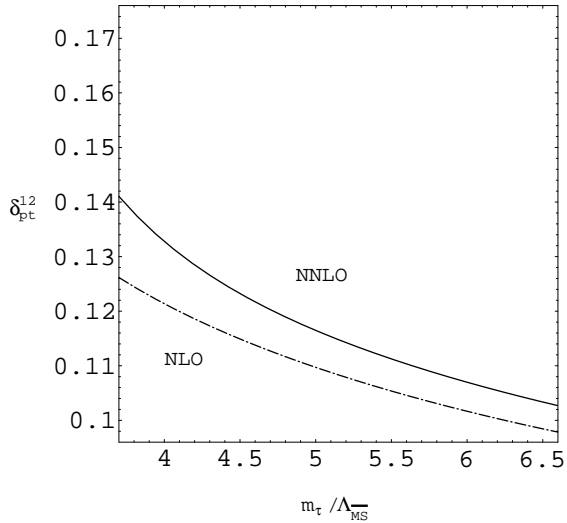,bbllx=202pt,bblly=295pt,bburx=410pt,bbury=497pt} 
\caption{$\delta_{pt}^{12}$ as a function of
$m_{\tau}/\Lambda_{\overline{MS}}^{(3)}$, obtained in the
$\overline{MS}$ scheme in NNLO (upper curve) and NLO (lower curve).}
\label{fig:msb}
\end{figure}

In figure~\ref{fig:endep} we show the NNLO PMS predictions for
$\delta_{pt}^{12}$  as a function of
$m_{\tau}/\Lambda_{\overline{MS}}^{(3)}$, together with the minimal
and maximal values obtained by varying the scheme parameters within the
region determined by the condition (\ref{allow}) with $l=2$.  We see
that the NNLO 
predictions for $\delta_{pt}^{12}$, obtained by numerically evaluating
the countour integral expression (\ref{eq:cont}), are free from
potentially dangerous RS instabilities, even for large values of
$\Lambda_{\overline{MS}}^{(3)}$. This situation is similar to that
encountered for $\delta_{pt}^{tot}$.  For comparison we also show the
PMS predictions obtained in the next-to-leading order (NLO).   We see
that RS dependence of the NNLO 
expression within the region defined by the condition (\ref{allow}) is
smaller than the difference between NNLO and NLO PMS predictions.
In a separate figure we show the NNLO and NLO predictions in the
$\overline{MS}$ scheme (figure~\ref{fig:msb}).

In order to see, how the optimization of the scheme parameters affects
the fits to the 
experimental data, we first test the accuracy of the approximation in
which one only retains the $O_{6}$ contribution in the SVZ
expansion. To this end we make a fit of $\alpha_{s}$ and $O_{6}$ in
the $\overline{MS}$ scheme and compare the results with the fit
performed by ALEPH~\cite{aleph95}, in which the $O_{4}$, $O_{6}$ and
$O_{8}$ contributions have been taken into account in the (1,0),
(1,1), (1,2) and (1,3) moments. To make the fits we use the following
expressions:
\be
R_{\tau}=3\times1.0194\,(0.991+\delta_{\tau}^{tot}-3.75\,O_{6}),
\label{eq:rtfit}
\ee
and 
\be
D_{\tau}^{12}=R_{\tau}^{12}/R_{\tau}^{00}=
\frac{13}{210}\frac{(1+\delta_{\tau}^{12}+
20.16\,O_{6})}{(1+\delta_{\tau}^{tot}-3.75\,O_{6})}.
\label{eq:d12}
\ee
If we take, following ALEPH~\cite{aleph95},
$R_{\tau}=3.645\pm0.024$ and 
$D_{\tau}^{12}=0.0570\pm0.0013$, we obtain
from the fit in the $\overline{MS}$ scheme
$\alpha_{s}(M_{Z}^{2})=0.1209\pm0.0013$ and $O_{6}=-0.0010\pm0.0012$.
This appears to be remarkably close to the values 0.121 and -0.0016
obtained in the full fit by ALEPH. This gives us confidence that the
``$O_{6}$~approximation'' captures the essential features of QCD
corrections in $\tau$ decays.

Let us now study how the RS dependence may affect the fit to the
experimental 
data. Let us take $R_{\tau}=3.635\pm0.016$, which is a weighted average
of three possible determinations~\cite{pdg96}, involving
$B_{e}=0.1783\pm0.0008$, $B_{\mu}=0.1735\pm0.0010$ and
$\tau_{\tau}=(291.0\pm1.5)\times10^{-15}\mbox{sec}$.  (These are the
so called ``our fit'' values. Using a set of ``our average'' values
given in~\cite{pdg96} we would obtain $R_{\tau}=3.643$.)  Let us also
take~\cite{cleo} $D_{\tau}^{12}=0.0559\pm0.0007$, which is the most
precise published value up to date.  Using the NNLO PMS expression we
obtain then $\alpha_{s}(M_{Z}^{2})=0.1188\pm0.0008$
($\alpha_{s}(m_{\tau}^{2})=0.330\pm0.008$) and
$O_{6}=-0.0021\pm0.0006$. Performing the same fit, but using now the
NNLO $\overline{MS}$ expression, we obtain
$\alpha_{s}(M_{Z}^{2})=0.1198$ and $O_{6}=-0.0020$. We see that the
value of the condensate practically does not change, but the value of
the strong coupling constant is reduced by amount approximately
equivalent to the experimental error.

It is also of some interest to repeat the fit using the NLO PMS
expression. We then obtain $\alpha_{s}(M_{Z}^{2})=0.1210$ and the same
value for $O_{6}$ as in NNLO. Taking the difference of the NNLO and
NLO PMS fits is 
perhaps the best way of estimating the accuracy of the perturbative
contribution in this problem, since we 
found
the QCD corrections to be quite stable with respect to change of the
scheme.   We see that thus obtained uncertainty of the fitted value of
$\alpha_{s}$ is of 
the order 0.0022, i.\ e.\ it is quite large, compared for example to
the experimental uncertainty. (This estimate of uncertainty involves only the
perturbative uncertainty --- to have  estimate of
theoretical uncertainty for the total QCD prediction one should also
consider the accuracy of the SVZ 
expansion itself.) It should be noted that the NNLO-NLO
difference is 
strongly RS dependent so it is essential to optimize the choice of the
scheme before comparing the predictions in successive orders.

Concluding, the renormalization scheme dependence of the perturbative
QCD corrections to $R_{\tau}$ and to the $R_{\tau}^{12}$ moment of the
invariant mass distribution in hadronic tau decays has been studied in
detail. The optimized predictions have been obtained using the
principle of minimal sensitivity as a guide to select the preferred
renormalization scheme. The stability of the predictions obtained via
the contour integral expression~(\ref{eq:cont}) has been verified, using a
specific condition to eliminate the schemes that have unnaturally
large expansion coefficients. However, the difference between
predictions in the conventionally used $\overline{MS}$ scheme and the
optimized predictions obtained in the PMS scheme was found to be
phenomenologically significant. Also, the difference between the NNLO
and NLO predictions in the PMS scheme was found to be significant,
indicating perhaps that the uncertainty in the perturbative expression
is larger than previously expected.

\end{document}